\documentclass[pdflatex,twocolumn,epjc3]{svjour3}          % twocolumn
\RequirePackage[T1]{fontenc}

\smartqed  % flush right qed marks, e.g. at end of proof

\usepackage{graphicx}
\usepackage{mathptmx}      % use Times fonts if available on your TeX system
\usepackage{flushend}
\usepackage[numbers,sort&compress]{natbib}
\usepackage[colorlinks,citecolor=blue,urlcolor=blue,linkcolor=blue]{hyperref}
\usepackage{epstopdf}
\usepackage{caption}
\usepackage{blindtext}
\journalname{Eur. Phys. J. C}

\begin{document}
\title{Is non-particle dark matter equation of state parameter evolving with time? }

%\subtitle{Do you have a subtitle?\\ If so, write it here}

\author{Souvik Ghose\thanksref{e1,addr1}
        \and
        Arunava Bhadra\thanksref{e2,addr1} %etc.
}

%\thankstext[$\star$]{t1}{Thanks to the title}
\thankstext{e1}{e-mail: dr.souvikghose@gmail.com}
\thankstext{e2}{e-mail: aru\_bhadra@yahoo.com}

\institute{High Energy $\&$ Cosmic Ray Research Center, University of North Bengal, Siliguri 734013, India\label{addr1}
%          \and
%          Second Address, Street, City, Country\label{addr2}
%          \and
%          \emph{Present Address:} Street, City, Country\label{addr3}
}

\date{Received:  / Accepted: date}
% The correct dates will be entered by the editor

\maketitle

\begin{abstract}
Recently, the so-called Hubble Tension, i.e. the mismatch between the local and the cosmological measurements of the Hubble parameter, has been  resolved when non-particle dark matter is considered which has a negative equation of state parameter ($\omega \approx -0.01$). We investigate if such a candidate can successfully describe the galactic flat rotation curves. It is found that the flat rotation curve feature puts a stringent constraint on the dark matter equation of state parameter $\omega$ and $\omega \approx -0.01$ is not consistent with flat rotational curves, observed around the galaxies. However, a dynamic $\omega$ of non-particle dark matter may overcome the Hubble tension without affecting the flat rotation curve feature.     
\end{abstract}

\section{Introduction}
\label{sec:intro}
The galactic rotation curve surveys \cite{de2008high} hint the presence of a dominating non-baryonic component of matter in galaxies that are non-luminous (dark) \cite{trimble1987,d2011dark}. Several other observations such as the cosmic microwave background (CMB) measurements \cite{de2000flat,komatsu2009five,collaboration2014planck,ade2015planck,ade2016}, baryon acoustic oscillations (BAO) \cite{eisenstein2005,anderson2012,slosar2013}, lensing in clusters \cite{allen2008,schrabback2010} also underpin the existence of dark matter. However, the nature of dark matter is still not known and it constitutes one of the unsolved mysteries in astrophysics at the present time.

There are several candidates for dark matter such as WIMPs \cite{jungman1996}, Axions \cite{duffy2009axions}, Sterile neutrinos \cite{boyarsky2019} etc. There are proposals for modifications at the fundamental theoretical level as well which include MOND \cite{milgrom1983, Mil2011} %bekenstein1984} 
 that suggests modifications in Newtonian dynamics. The conformal gravitational theory, which is based on Weyl symmetry, can also explain flat rotation curves of galaxies without the need for dark matter \cite{mannheim1989,tang2020weyl}. The evidence of non-baryonic dark matter from the CMB data, however, questions the MOND like schemes.  

Recently, Popławski \cite{poplawski2019} has argued that the discrepancy on the value of Hubble parameter obtained from the cosmic microwave background \cite{Planck2018,hinshaw13} radiation and that obtained from the measurements using Cepheid variable stars \cite{riess2021} can be avoided if dark matter is considered as non-particle with the negative equation of state parameter. Such a proposal is also interesting because of the non-detection of any dark matter particles in any direction (laboratory) observations so far despite a significant enhancement of experimental sensitivity over the last decade or so. Hence the non-particle dark matter hypothesis deserves further investigation.

In this work, we would like to examine whether the non-particle dark matter could describe the rotation curve feature of spiral galaxies consistently. We shall particularly consider two velocity profiles, the universal velocity profile proposed by Persic \emph{et al.} \cite{persic1996} and secondly the exact flat nature of galactic rotation curve at galactic halo region which essentially implies that the tangential velocity is constant, independent of radial distance ($r$) for large r. Note that the gravitational field of galactic halo considering the different velocity profiles in the halo region has already been studied in the literature \cite{rah10}, \cite{bar15} but the negative value of the equation of state parameter ($\omega$) was never considered as it sounds unrealistic. Here our objective is to extend the analysis of \cite{rah10}, \cite{bar15} for negative $\omega$. %Besides, we shall also consider the velocity profile of Persic et al. \cite{Persic1996} along with a barotropic dark fluid, $p=w\rho$, where p is the pressure, $\rho$ is the energy density and w is the equation of state parameter, and 
Subsequently, we shall explore whether small negative $\omega$, as needed to consistently match the Hubble constants estimated from the observations of two distance scales, is consistent with the observed galactic flat rotation curve feature.        
  
The organization of the paper is the following. In the next section (sec. \ref{sec:nonp} we shall briefly discuss the non-particle dark matter. In section \ref{sec:ghal} we shall obtain space-time geometry of galactic halo considering the observed flat rotation curve feature of galaxies in presence of non-particle dark matter. Subsequently we evaluate the expressions for pressure and energy density of non-particle dark matter field and study the viable parametric space for the parameter w in section 4. In the same section, we shall exploit velocity vs baryonic mass data of a large sample of galaxies to correlate one of the parameters of the solution with the baryonic mass of the galaxy. We discuss our results in section \ref{sec:disc} and conclude in the same section. 

\section{Non-particle dark matter}  
\label{sec:nonp}
The Hubble parameter H ($\equiv \frac{\dot{a}}{a}$, where over dot denotes derivative with respect to time) essentially reflects the expansion rate of the Universe. The present day value of Hubble parameter ($H_o$) is estimated from various observations involving different distance scales. The direct measurements in the local universe particularly from separations to supernova type 1a employing Cepheids, masers etc as distant ladders %standard candles 
gives $H_o = 100 h km/s/Mpc$ where $h = 0.732 \pm 0.013$ \cite{riess2021} %$h=0.742 \pm 0.018$ \cite{riess2019}. 
On the other hand the precision observations of the Cosmic Microwave Background (CMB) and large scale structure suggests a smaller value. Based on the concordance $\Lambda$CDM (cosmological constant and cold dark matter) model the Planck data reveals $h = 0.6736 \pm 0.0054$ \cite{Planck2018}. A large number of measurements with different techniques are available for $H_o$. The median value of 331 such measurements was found $h = 0.674 \pm 0.05$ \cite{gott2001} which slightly increases to $h = 0.68 \pm 0.028 $ when total 553 measurements was considered a decade later \cite{Chen2011}. Note that the error quoted in \cite{gott2001} is based on 2$\sigma$ confidence level whereas the other errors quoted above are 1$\sigma$ error.   

Several suggestions have been put forward in the literature for the origin of the stated discrepancy in the value of the Hubble parameter. Popławski, in his recent paper \cite{poplawski2019}, has argued that the discrepancy in the value of Hubble parameter as obtained from cosmological and local observations can be settled by considering a non-particle dark matter with an equation of state $p=\omega \rho$ and $\omega \approx -0.01$. The idea of Popławski rests on the fact that one cannot directly estimate the dark matter content ($\Omega_{DM}$) in the Universe from the CMB data but rather judges $\Omega_{DM} h^2 (1+z)^3$. When $\Omega_{DM}$ is known from some other measurements, one can infer $h$ from the CMB data. However, if the dark matter equation of state is described as $\rho_{DM} = \omega p_{DM}$, the time evolution of dark matter content will be modified and consequently one obtains $\Omega_{DM} h^2 (1+z)^{3(1+\omega)}$ from the CMB data instead of $\Omega_{DM} h^2 (1+z)^3$ as often used. Demanding that the h value from CMB data must match with that obtained from local observations Popławski found that such a bid satisfies when $\omega \approx -0.01$. However, observation of flat rotation curves has been one of the motivations behind consideration of the dark matter in galaxies and it is, therefore, prudent to further investigate if such a value of $\omega$ would describe the flat rotation curves.

\section{Gravitational field of galactic halo and non-particle dark matter}
\label{sec:ghal}

Observations reveal that the orbits of stars and interstellar gas around the center of a spiral galaxy are approximately circular. %The mass distributions in galactic bulge and halo are at least approximately spherically symmetric which dominate the total mass in the galaxy. 
The gravitational field in the halo region of a spiral galaxy thus should  well be approximated as spherical symmetric. The general spherically symmetric static line element is given by     

\begin{equation}
\label{eq:metric}
ds^{2}=-e^{\mu(r)}dt^{2}+e^{\nu(r)}dr^{2}+r^{2}(d\theta^{2}+sin^{2}\theta
d\phi^{2}),
\end{equation}

where $\mu(r)$ and $\nu(r)$ are functions of radial distance r only.  For the above metric the Einstein field equations $R_{mu\nu} = 8\pi G T_{\mu\nu}$ give ($c=1$ through out the paper):

\begin{equation}
\label{eq:efe1}
e^{-\nu}\left[  \frac{\nu^{\prime}}{r}-\frac{1}{r^{2}}\right]
+\frac{1}{r^{2}}=8\pi G\rho
\end{equation}

\begin{equation}
\label{eq:efe2}
e^{-\nu}\left[  \frac{1}{r^{2}}+\frac{\mu^{\prime}}{r}\right]  -\frac
{1}{r^{2}}=8\pi Gp
\end{equation}

\begin{equation}
\label{eq:efe3}
\frac{1}{2}e^{-\nu}\left[  \frac{1}{2}(\mu^{\prime})^{2}+\mu^{\prime
\prime}-\frac{1}{2}\nu^{\prime}\mu^{\prime}+\frac{1}{r}({\mu^{\prime
}-\nu^{\prime}})\right]  =8\pi Gp.
\end{equation}

%We shall now exploit the observed flat nature of rotation curves of galaxies to get a solution of the metric coefficients for a matter field with the equation of state $p=w\rho$.   
In the above we have taken the stress-energy tensor of dark matter as $T_{\mu\nu}=\left(\rho + p\right)U_{\mu}U_{\nu} + pg_{\mu\nu}$, $\rho$ is the rest frame energy density, $p$ is the pressure and $U$ is the four velocity, as considered by Poplawski for non-particle dark matter \cite{poplawski2019}. Note that though the above consideration of stress energy tensor resembles with that of ideal fluid but the fluid-mechanical interpretation is not proper in view of non-particle feature (like cosmological constant) of dark matter.    

Restricting geodesic motion in the equatorial plane, the tangential velocity ($v_{\phi}$) of a test particle in a circular orbit in the space-time metric of Eq.(1) is given by

\begin{equation}
\label{eq:cfrcv}
v_{\phi}^2=\frac{r(e^{-\mu(r)})_r}{2e^{-\mu(r)}},
\end{equation}
where subscript r refers to derivative with respect to r. 

\subsection{The metric coefficients for the Universal velocity profile}

Because of the overall resemblance of natures of rotation curves for spiral galaxies a number of attempts have been made to describe the rotation curves by a single mathematical formula. From a sample of around 1100 optical and radio rotation curves, Persic et al. \cite{persic1996} expressed the rotation curves, comprising both disk and halo components, as a function of total luminosity and radius by a simple formula as given below

\begin{equation}
\label{eq:p1}
v_{\phi}^2=v_{o}^2 \frac{r^2}{r^2+r_o^2}
\end{equation}

where $v_o$ and $r_o$ are two constants, $v_o$ is a function of luminosity of the galaxy. The above expression does not contain any free parameters, and it is often called as universal rotation curve of galaxies.   

For the universal velocity profile (Eq.(\ref{eq:p1})) the Eq.(\ref{eq:cfrcv}) gives     

\begin{equation}
\label{eq:p2}
\mu(r) \simeq v_o^2 ln [(r^2+r_o^2)/R^2] 
\end{equation}  
where R is a constant. %One can recover Eq.(\ref{eq:cfrc}) from the above equation when $r>>r_o$. 

The solution of $e^{\nu}$ obtained by inserting the above expression in (\ref{eq:efe2}), and (\ref{eq:efe3}) is quite clumsy, involving hypergeometric function and is difficult to handle. Barranco et al. \cite{bar15}, instead considered the Newtonian limit (i.e. the pressure much smaller than the density) of Eqs. (\ref{eq:efe2}), and (\ref{eq:efe3}) and subsequently found    

\begin{equation}
\label{eq:p3}
e^{-\nu}= 1- 2v_o^2 \frac{r^2}{r^2+r_o^2}
\end{equation}  
where k and R are constants.

The Eqs.(\ref{eq:efe2}) and (\ref{eq:efe3}) give the expressions for energy density and pressure

\begin{equation}
\label{eq:p4}
\rho= \frac{v_o^2}{4 \pi}  \frac{r^2+3r_o^2}{(r^2+r_o^2)^2} 
\end{equation}  

\begin{equation}
\label{eq:p5}
p= \frac{v_o^4}{8 \pi}  \frac{r^2+2r_o^2}{(r^2+r_o^2)^2} 
\end{equation}  

The central density and pressure (at $r=0$) are given by

\begin{equation}
\label{eq:p7}
\rho_o = \frac{3 v_o^2}{4\pi r_o^2}
\end{equation}  
and
\begin{equation}
\label{eq:p8}
p_o = \frac{v_o^4}{4\pi r_o^2}
\end{equation}  
 
The equation of state of the matter field leading to the solutions (\ref{eq:p2}) and (\ref{eq:p3}) can be expressed as \cite{bar15}

\begin{equation}
\label{eq:p6}
p= p_o \left(\frac{3 \rho}{4 \rho_o} - \frac{1}{16} \left(1-\sqrt{1+24 \left(\frac{\rho}{\rho_o}\right)} \right) \right) 
\end{equation}  

%The parameters $\rho_o$ and $p_o$ are essentially the density and pressure at  r=0.    
Under the condition that $\rho << \rho_o$, the equation of state can be expressed as $p= \omega \rho$ where the equation of state parameter is given by 
\begin{equation}
\label{eq:p9}
\omega= \frac{3 p_o}{2 \rho_o}  
\end{equation}

Since both $\rho_o$ and $p_o$ are positive, w has to be positive which implies that non-particle dark matter cannot reproduce the universal  rotation curves of galaxies. However, since Newtonian limit of Einstein field equations was considered in deriving the equation of state of the matter field that leads to universal rotation curve we cannot rule out the possibility of negative $\omega$ in full relativistic treatment.   

\subsection{The metric coefficients for exactly flat rotation curve feature}
When $r>>r_o$ the universal velocity profile (Eq.(\ref{eq:p1}) reduces to a constant $v_{\phi}$, independent of radial distance. The observation of a large number of galactic rotation curves exhibit such almost constant $v_{\phi}$ in the galactic halo region. For constant $v_{\phi}$ the Eq. (\ref{eq:cfrcv}) straightway gives    

\begin{equation}
\label{eq:cfrc}
e^{\mu(r)}= B_0 r^q,
\end{equation}

where $q \equiv 2(v_{\phi})^{2}$ and $B_0$ is an integration constant. %The observed rotational curve profile in the region dominated by dark matter is such that the rotational velocity $v^{\phi}$ becomes approximately a constant with $v^{\phi}\sim10^{-3}$ ( 300km/s) for a typical galaxy. 
Inserting the solution of $e^{\mu(r)}$ (\ref{eq:cfrc}) in the Eqs. (\ref{eq:efe2}) and (\ref{eq:efe3}) one gets  %

\begin{equation}
\label{eq:efe4}
(e^{-\nu})_r +\frac{ae^{-\nu}}{r}=\frac{b}{r},
\end{equation}
where%

\begin{equation}
\label{eq:efea}
a=-\frac{4(1+q)-q^{2}}{2+q}%
\end{equation}
and%

\begin{equation}
\label{eq:efec}
b=-\frac{4}{2+q}.
\end{equation}

The exact solution of Eq.(\ref{eq:efe4}) reads
\begin{equation}
\label{eq:efe4sol}
e^{-\nu}=\frac{b}{a}+\frac{D}{r^{a}},
\end{equation}
where $D$ is an integration constant. The Eq.(\ref{eq:metric}) along with Eqs. (\ref{eq:cfrc}) and (\ref{eq:efe4sol}) give the space-time geometry of galactic halo for for arbitrary matter field with isotropic pressure field.  

Inserting the expressions for $\mu(r)$ and $\nu(r)$ in Einstein field equations ((\ref{eq:efe2}) - (\ref{eq:efe3}), we get the expressions for $\rho$, $p$ as%

\begin{equation}
\label{eq:exprho}
\rho= \frac{1}{8 \pi G } \left[  \frac{q(4-q)}{4+4q-q^{2}} r^{-2} -
\frac{D(6-q)(1+q)}{2+q} r^{q(2-q)/(2+q)}\right]
\end{equation}

\begin{equation}
\label{eq:expp}
p=\frac{1}{8\pi G}\left[  \frac{q^{2}}{4+4q-q^{2}}r^{-2}%
+D(1+q)r^{q(2-q)/(2+q)}\right]
\end{equation}
 
Here an analytical expression of equation of state is not available. 
%\section{The viable parametric space for w of non-particle dark matter}

%As mentioned in section 2 the non-particle dark matter is described by a barotropic dark fluid with equation of state $p=w\rho$ and is characterized by negative equation of state parameter w. Below we shall study the equation of state of the dark matter in galactic halo region. 
%explore whether it is possible to relate p with $\rho$ by $p=w\rho$ with negative under some conditions    
%In general the equation of state for the dark matter in galactic halo as obtained above cannot be expressed by simple $p=w\rho$ with w as a constant. 

%\subsection{Exactly flat velocity profile}

The ratio ($\omega$ of p to $\rho$ for the dark matter field is obtained from Eqs.(\ref{eq:exprho}) and (\ref{eq:expp}), %

\begin{equation}
\label{eq:w}
\omega=\frac{p}{\rho}=\frac{q^{2}r^{a}+D(1+q)(4+4q-q^{2})}{q(4-q)r^{a}%
-D(6-q)(1+q)(4+4q-q^{2})/(2+q)}. 
\end{equation}
Since $q$ is small, $\omega$ can be written as: 
\begin{equation}
\label{eq:omsim}
\omega\simeq\frac{q^{2}r^{a}+4D}{4qr^{a}-12D}.
\end{equation}

It appears from the above equation that $\omega$, in general, is a function of radial distance. If we demand that $\omega$ has to be a constant that can be achieved when either $q^2 r^a >>D$ or $D>> q r^a$. The former condition, however, contradicts the condition of negativity of $\omega$. So the only option remains is that $D>> q r^a$ which gives $\omega \sim - 1/3$. 

\begin{figure}
\centering
\includegraphics[scale=0.4]{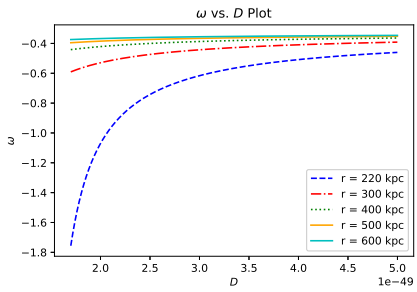}
\caption{Variaion of $\omega$ with $D$ for different galaxy redii ($r$)}
\label{fig:omd}
\end{figure}

%The non-particle dark matter is characterized by negative w. Such a condition is achievable if D is positive and $D>qr^a/3$ or if D is negative and $|D| > q^2 r^a/4$. Now if we assume that the galactic halo is filled with a dark matter fluid of fixed equation of state parameter w then the above equation implies that either $q^2 r^a >>D$ or $D>> q r^a$. The former condition, however, contradicts the condition of negativity of w. So the only option remains is that $D>> q r^a$ which gives $w \sim - 1/3$. 
To exemplify the above-mentioned point without restricting $\omega$ as independent of r we have plotted $\omega \; vs. \; D$ for different $r$ values from eq. (\ref{eq:omsim}) for a typical rotation speed of $200 $ $km/s$. The fig. (\ref{fig:omd}) shows a few of such plots. We note that negative values of $\omega$ are consistent with observations as long as $\omega < -.33$ which interestingly resemblance with the dark energy condition. Clearly, $\omega$ value never reaches $-0.01$ as found by Popławski \cite{poplawski2019} to obtain a consistent unique Hubble constant from the local and cosmological observations.

%\subsection{radial distance dependent tangential velocity}
%\section{.....}
%\label{anal}

%\subsection{Flat rotation curve and dark matter equation of state}
%\label{flatom}

\section{Discussion and Conclusion}
\label{sec:disc}
In the Poplawski model the equation of state of dark matter is assumed that of an ideal fluid $p = w \rho$. In the present work we have employed exactly the same assumption and explored whether the static spherically symmetric space-time of galactic halo is consistent with such an assumption of equation of state with negative w. As mentioned in \cite{poplawski2019} a conventional fluid with w < 0 is unstable that likely to lead inhomogeneity. The instability of the fluid might affect the stability of the static solution of the halo space-time which we have ignored in the present work. We have also not considered the issue of inhomogeneity in dark matter distribution. We focus on the non-particle nature of dark matter without considering the fluid mechanical interpretation of dark matter equation of state.  

The present study suggests that it is not enough to resolve the tension between the current expansion rate from the global and local measurements for a consistent description of dark matter but the analysis of galaxy rotation curves also needs to be included. In particular Popławski \cite{poplawski2019} has floated an interesting idea of non-particle dark matter to overcome the discrepancy in the measured values of the Hubble constant from the local and cosmological observations. However, here we find that such a scheme is not compatible with the observed feature of the galactic rotation curves, at least as long as inhomogeneity of dark matter distribution and stability issues are not taken into consideration. The magnitude of the equation of state parameter required to remove the discrepancy in the estimation of Hubble parameter is found smaller than that admissible for matching the galactic rotation curve.

The admissible value of negative $\omega$ ($ \sim - 1/3$) as allowed from the flat galactic rotation curve feature resembles the dark energy equation of state parameter required for the late time acceleration. It is not clear at this stage that whether such similarity is just a simple coincidence or it indicates some deeper correlation between dark matter and dark energy. The admissible negative $\omega$ value, however, leads to a much larger and unrealistic $h$ value if one follows the procedure of Popławski \cite{poplawski2019}. Moreover, an analysis of CMBR data as observed by Planck together with the baryonic oscillation data (BAO) does not support such large negative $\omega$ \cite{kopp2018}.
%It thus appears that the tension between the current expansion rate from the global and local measurements may have some different origin.      

Does $\omega$ of non-particle evolve with time? If it evolves from its value $\approx -0.01$ at the recombination epoch to a vanishing or a slightly positive $\omega$ or even towards $\omega<-0.33$ at present epoch, the Hubble tension will be resolved, as shown by Popławski \cite{poplawski2019}, without compromising with the galactic flat rotation curve feature.  Several models have been proposed in cosmology considering the evolving equation of state for dark energy. For dark matter, such a feature is not very common. However, since non-particle dark matter can be described by a scalar field (or by some other field) the evolution of $\omega$ seems quite feasible theoretically. Even it may be possible to choose the time variation of $\omega$ in such a way that the non-particle dark matter also contributes to structure formation %as required for the galaxy formation at early stages of cosmic history, 
which is not possible in the constant $\omega$ scheme.

\section*{Acknowledgments}
The authors thank an anonymous reviewer for useful comments that help us to improve the manuscript. The work  of AB is supported by Science and Engineering Research Board (SERB), DST through grant number CRG/2019/004944.
%\end*{Acknowledgments}


\begin{thebibliography}{10}

\providecommand{\url}[1]{{#1}}
\providecommand{\urlprefix}{URL }
\expandafter\ifx\csname urlstyle\endcsname\relax
  \providecommand{\doi}[1]{DOI \discretionary{}{}{}#1}\else
  \providecommand{\doi}{DOI \discretionary{}{}{}\begingroup
  \urlstyle{rm}\Url}\fi

\bibitem{de2008high}
W.~De~Blok, F.~Walter, E.~Brinks, C.~Trachternach, S.~Oh, R.~Kennicutt~Jr, The
  Astronomical Journal \textbf{136}(6), 2648 (2008)

\bibitem{trimble1987}
V.~Trimble, Annual review of astronomy and astrophysics \textbf{25}(1), 425
  (1987)

\bibitem{d2011dark}
G.~D'Amico, M.~Kamionkowski, K.~Sigurdson, in \emph{Dark Matter and Dark
  Energy} (Springer, 2011), pp. 241--272

\bibitem{de2000flat}
P.~de~Bernardis, P.A. Ade, J.~Bock, J.~Bond, J.~Borrill, A.~Boscaleri,
  K.~Coble, B.~Crill, G.~De~Gasperis, P.~Farese, et~al., Nature
  \textbf{404}(6781), 955 (2000)

\bibitem{komatsu2009five}
E.~Komatsu, J.~Dunkley, M.~Nolta, C.~Bennett, B.~Gold, G.~Hinshaw, N.~Jarosik,
  D.~Larson, M.~Limon, L.~Page, et~al., The Astrophysical Journal Supplement
  Series \textbf{180}(2), 330 (2009)

\bibitem{collaboration2014planck}
P.~Collaboration, P.~Ade, N.~Aghanim, C.~Armitage-Caplan, M.~Arnaud,
  M.~Ashdown, F.~Atrio-Barandela, J.~Aumont, C.~Baccigalupi, A.~Banday, et~al.,
  Astron. Astrophys \textbf{571}, A16 (2014)

\bibitem{ade2015planck}
P.A. Ade, N.~Aghanim, D.~Alina, M.~Alves, C.~Armitage-Caplan, M.~Arnaud,
  D.~Arzoumanian, M.~Ashdown, F.~Atrio-Barandela, J.~Aumont, et~al., Astron.
  Astrophys. \textbf{576}, A104 (2015)

\bibitem{ade2016}
{Planck Collaboration}, P.A.R. {Ade}, N.~{Aghanim}, H.D. {Aller}, M.F. {Aller},
  M.~{Arnaud}, J.~{Aumont}, C.~{Baccigalupi}, A.J. {Banday}, R.B. {Barreiro},
  N.~{Bartolo}, E.~{Battaner}, K.~{Benabed}, A.~{Benoit-L{\'e}vy}, J.P.
  {Bernard}, M.~{Bersanelli}, P.~{Bielewicz}, A.~{Bonaldi}, L.~{Bonavera}, J.R.
  {Bond}, J.~{Borrill}, F.R. {Bouchet}, C.~{Burigana}, E.~{Calabrese},
  A.~{Catalano}, H.C. {Chiang}, P.R. {Christensen}, D.L. {Clements}, L.P.L.
  {Colombo}, F.~{Couchot}, B.P. {Crill}, A.~{Curto}, F.~{Cuttaia}, L.~{Danese},
  R.D. {Davies}, R.J. {Davis}, P.~{de Bernardis}, A.~{de Rosa}, G.~{de Zotti},
  J.~{Delabrouille}, C.~{Dickinson}, J.M. {Diego}, H.~{Dole}, S.~{Donzelli},
  O.~{Dor{\'e}}, A.~{Ducout}, X.~{Dupac}, G.~{Efstathiou}, F.~{Elsner}, H.K.
  {Eriksen}, F.~{Finelli}, O.~{Forni}, M.~{Frailis}, A.A. {Fraisse},
  E.~{Franceschi}, S.~{Galeotta}, S.~{Galli}, K.~{Ganga}, M.~{Giard},
  Y.~{Giraud-H{\'e}raud}, E.~{Gjerl{\o}w}, J.~{Gonz{\'a}lez-Nuevo}, K.M.
  {G{\'o}rski}, A.~{Gruppuso}, M.A. {Gurwell}, F.K. {Hansen}, D.L. {Harrison},
  S.~{Henrot-Versill{\'e}}, C.~{Hern{\'a}ndez-Monteagudo}, S.R. {Hildebrandt},
  M.~{Hobson}, A.~{Hornstrup}, T.~{Hovatta}, W.~{Hovest}, K.M. {Huffenberger},
  G.~{Hurier}, A.H. {Jaffe}, T.R. {Jaffe}, E.~{J{\"a}rvel{\"a}},
  E.~{Keih{\"a}nen}, R.~{Keskitalo}, T.S. {Kisner}, R.~{Kneissl}, J.~{Knoche},
  M.~{Kunz}, H.~{Kurki-Suonio}, A.~{L{\"a}hteenm{\"a}ki}, J.M. {Lamarre},
  A.~{Lasenby}, M.~{Lattanzi}, C.R. {Lawrence}, R.~{Leonardi}, F.~{Levrier},
  M.~{Liguori}, P.B. {Lilje}, M.~{Linden-V{\o}rnle}, M.~{L{\'o}pez-Caniego},
  P.M. {Lubin}, J.F. {Mac{\'\i}as-P{\'e}rez}, B.~{Maffei}, D.~{Maino},
  N.~{Mandolesi}, M.~{Maris}, P.G. {Martin}, E.~{Mart{\'\i}nez-Gonz{\'a}lez},
  S.~{Masi}, S.~{Matarrese}, W.~{Max-Moerbeck}, P.R. {Meinhold},
  A.~{Melchiorri}, A.~{Mennella}, M.~{Migliaccio}, M.~{Mingaliev}, M.A.
  {Miville-Desch{\^e}nes}, A.~{Moneti}, L.~{Montier}, G.~{Morgante},
  D.~{Mortlock}, D.~{Munshi}, J.A. {Murphy}, F.~{Nati}, P.~{Natoli},
  E.~{Nieppola}, F.~{Noviello}, D.~{Novikov}, I.~{Novikov}, L.~{Pagano},
  F.~{Pajot}, D.~{Paoletti}, B.~{Partridge}, F.~{Pasian}, T.J. {Pearson},
  O.~{Perdereau}, L.~{Perotto}, V.~{Pettorino}, F.~{Piacentini}, M.~{Piat},
  E.~{Pierpaoli}, S.~{Plaszczynski}, E.~{Pointecouteau}, G.~{Polenta}, G.W.
  {Pratt}, V.~{Ramakrishnan}, E.A. {Rastorgueva-Foi}, A.C. {S Readhead},
  M.~{Reinecke}, M.~{Remazeilles}, C.~{Renault}, A.~{Renzi}, J.L. {Richards},
  I.~{Ristorcelli}, G.~{Rocha}, M.~{Rossetti}, G.~{Roudier}, J.A.
  {Rubi{\~n}o-Mart{\'\i}n}, B.~{Rusholme}, M.~{Sandri}, M.~{Savelainen},
  G.~{Savini}, D.~{Scott}, Y.~{Sotnikova}, V.~{Stolyarov}, R.~{Sunyaev},
  D.~{Sutton}, A.S. {Suur-Uski}, J.F. {Sygnet}, J.~{Tammi}, J.A. {Tauber},
  L.~{Terenzi}, L.~{Toffolatti}, M.~{Tomasi}, M.~{Tornikoski}, M.~{Tristram},
  M.~{Tucci}, M.~{T{\"u}rler}, L.~{Valenziano}, J.~{Valiviita}, E.~{Valtaoja},
  B.~{Van Tent}, P.~{Vielva}, F.~{Villa}, L.A. {Wade}, A.E. {Wehrle}, I.K.
  {Wehus}, D.~{Yvon}, A.~{Zacchei}, A.~{Zonca}, Astron. Astrophys.
  \textbf{596}, A106 (2016).
\newblock \doi{10.1051/0004-6361/201527780}

\bibitem{eisenstein2005}
D.J. Eisenstein, I.~Zehavi, D.W. Hogg, R.~Scoccimarro, M.R. Blanton, R.C.
  Nichol, R.~Scranton, H.J. Seo, M.~Tegmark, Z.~Zheng, et~al., The
  Astrophysical Journal \textbf{633}(2), 560 (2005)

\bibitem{anderson2012}
L.~Anderson, E.~Aubourg, S.~Bailey, D.~Bizyaev, M.~Blanton, A.S. Bolton,
  J.~Brinkmann, J.R. Brownstein, A.~Burden, A.J. Cuesta, et~al., Monthly
  Notices of the Royal Astronomical Society \textbf{427}(4), 3435 (2012)

\bibitem{slosar2013}
A.~Slosar, V.~Ir{\v{s}}i{\v{c}}, D.~Kirkby, S.~Bailey, T.~Delubac, J.~Rich,
  {\'E}.~Aubourg, J.E. Bautista, V.~Bhardwaj, M.~Blomqvist, et~al., Journal of
  Cosmology and Astroparticle Physics \textbf{2013}(04), 026 (2013)

\bibitem{allen2008}
S.~Allen, D.~Rapetti, R.~Schmidt, H.~Ebeling, R.~Morris, A.~Fabian, Monthly
  Notices of the Royal Astronomical Society \textbf{383}(3), 879 (2008)

\bibitem{schrabback2010}
T.~Schrabback, J.~Hartlap, B.~Joachimi, M.~Kilbinger, P.~Simon, K.~Benabed,
  M.~Brada{\v{c}}, T.~Eifler, T.~Erben, C.D. Fassnacht, et~al., Astron.
  Astrophys. \textbf{516}, A63 (2010)

\bibitem{jungman1996}
G.~Jungman, M.~Kamionkowski, K.~Griest, Physics Reports \textbf{267}(5-6), 195
  (1996)

\bibitem{duffy2009axions}
L.D. Duffy, K.~Van~Bibber, New Journal of Physics \textbf{11}(10), 105008
  (2009)

\bibitem{boyarsky2019}
A.~Boyarsky, M.~Drewes, T.~Lasserre, S.~Mertens, O.~Ruchayskiy, Progress in
  Particle and Nuclear Physics \textbf{104}, 1 (2019)

\bibitem{milgrom1983}
M.~Milgrom, The Astrophysical Journal \textbf{270}, 365 (1983)

%\bibitem{bekenstein1984}
%J.~Bekenstein, M.~Milgrom, The Astrophysical Journal \textbf{286}, 7 (1984)

\bibitem{Mil2011}
M. Milgrom, Acta Phys. Pol. B \textbf{42}, 2175 (2011)

\bibitem{mannheim1989}
P.D. Mannheim, D.~Kazanas, The Astrophysical Journal \textbf{342}, 635 (1989)

\bibitem{tang2020weyl}
Y.~Tang, Y.L. Wu, Physics Letters B p. 135320 (2020)

\bibitem{poplawski2019}
N.J. Pop{\l}awski, The European Physical Journal C \textbf{79}(9), 734 (2019)

\bibitem{hinshaw13}
G.~{Hinshaw}, D.~{Larson}, E.~{Komatsu}, D.N. {Spergel}, C.L. {Bennett},
  J.~{Dunkley}, M.R. {Nolta}, M.~{Halpern}, R.S. {Hill}, N.~{Odegard},
  L.~{Page}, K.M. {Smith}, J.L. {Weiland}, B.~{Gold}, N.~{Jarosik}, A.~{Kogut},
  M.~{Limon}, S.S. {Meyer}, G.S. {Tucker}, E.~{Wollack}, E.L. {Wright}, The
  Astrophysical Journals \textbf{208}(2), 19 (2013).
\newblock \doi{10.1088/0067-0049/208/2/19}

\bibitem{riess2021}
A. G. Riess, S.~Casertano, W.~Yuan, J. B. Bowers, L.M. Macri, J. C. Zinn, D.~Scolnic, Astrophys. J. Lett. 908 L6 (2021)

%\bibitem{riess2019}
%A.G. {Riess}, S.~{Casertano}, W.~{Yuan}, L.M. {Macri}, D.~{Scolnic}, The Astrophysical Journal \textbf{876}(1), 85 (2019). \newblock \doi{10.3847/1538-4357/ab1422}

\bibitem{persic1996}
M.~{Persic}, P.~{Salucci}, F.~{Stel}, Mon. Not. R Astron. Soc. \textbf{281}(1),
  27 (1996).
\newblock \doi{10.1093/mnras/278.1.27}

\bibitem{rah10}
F.~{Rahaman}, K.K. {Nandi}, A.~{Bhadra}, M.~{Kalam}, K.~{Chakraborty}, Physics
  Letters B \textbf{694}(1), 10 (2010).
\newblock \doi{10.1016/j.physletb.2010.09.038}

\bibitem{bar15}
J.~{Barranco}, A.~{Bernal}, D.~{N{\'u}{\~n}ez}, Mon. Not. R Astron. Soc.
  \textbf{449}(1), 403 (2015).
\newblock \doi{10.1093/mnras/stv302}

\bibitem{Planck2018} Planck Collaboration, N. Aghanim et al., 
Astron. Astrophys 641, A6 (2020) (ArXiv 1807.06209)
\newblock \doi{10.1051/0004-6361/201833910}

\bibitem{gott2001}
I.~{Gott}, J.~Richard, M.S. {Vogeley}, S.~{Podariu}, B.~{Ratra}, The   Astrophysical Journal \textbf{549}(1), 1 (2001).\newblock \doi{10.1086/319055}

\bibitem{Chen2011}
G, Chen and B. Ratra  PASP 123 1127 (2011)  arXiv 1105.5206

\bibitem{kopp2018}
M. Kopp, C. Skordis, D.B. Thomas, S. Ilic, Phys. Rev. Lett. 120,
221102 (2018)

\end{thebibliography}
\end{document}